\documentclass[11pt]{article}
\usepackage{graphicx,fancyhdr}
\usepackage{amsmath,amsthm, amssymb, latexsym}
\usepackage[square,sort,comma,numbers]{natbib} 
\usepackage[T1]{fontenc}
\title{From evaluation to learning: Some aspects of designing a cyber-university}

\author{\Large Anna Helga Jonsdottir, Gunnar Stefansson}
\begin{document}

\maketitle

\begin{center}
{\bf Abstract}
\end{center}

Research is described on a system for web-assisted education and how
it is used to deliver on-line drill questions, automatically suited to
individual students. The system can store and display all of the
various pieces of information used in a class-room (slides, examples,
handouts, drill items) and give individualized drills to participating
students. The system is built on the basic theme that it is for learning rather
than evaluation. 

Experimental results shown here imply that both the item database and
the item allocation methods are important and examples are given on
how these need to be tuned for each course. Different item allocation
methods are discussed and a method is proposed for comparing several
such schemes. It is shown that students improve their knowledge
while using the system. Classical statistical models which do not
include learning, but are designed for mere evaluation, are therefore
not applicable.

A corollary of the openness and emphasis on learning is that the
student is permitted to continue requesting drill items until the
system reports a grade which is satisfactory to the student. An
obvious resulting challenge is how such a grade should be computed so
as to reflect actual knowledge at the time of computation, entice the
student to continue and simultaneously be a clear indication for the
student.  To name a few methods, a grade can in principle be computed
based on all available answers on a topic, on the last few answers or
on answers up to a given number of attempts, but all of these have
obvious problems.

\newpage

\section{Background}
\subsection{This project}
This R\&D project in web-assisted education attempts to address the following issues:
\begin{itemize}
\item a shortage of experienced educators in mathematics and statistics
\item a lack of implemented standards for education (baseline outputs)
  in mathematics and statistics 
\item a lack of applied statistics courses for researchers and students in other fields.
\end{itemize}
The approach taken in the project includes the following components:
\begin{itemize}
\item design freely available web-based material to degrees in mathematics and applied statistics
\item allocate personalized drill items using an IAA (item allocation algorithm)
\item invoke the student's incentive for a high grade using a GS (grading scheme)
\end{itemize}

The primary research question addressed by the project is the search
for the best \textbf{ item allocation algorithm} or how one can best
select the next drill item for the student, with the
\textbf{grading scheme} a close second.

Many systems are available to instruct on specific topics and
considerable research has been conducted on how to fine-tune
presentation of material or drills on specific topics.  The system to
be designed is \textbf{generic}, however, uses mostly multiple-choice
drill items, but delivered in a very specialized manner, and can be
tuned to any field of interest.

\subsection{Classical testing methods}
The field of computerized testing has been around for a number of
decades.  Item response theory \cite{lord1980applications} has been
used to design, analyse, and grade computerized tests of abilities.
Data are binary responses to each item and such data are commonly
analysed with logistic regression models. The three-parameter logistic
model is often used to link the probability of a correct answer to the
students ability
\begin{equation}
\label{eqn:3pl}
P_{si} = P(Y_{si}=1|\theta_s;a_i,b_i,c_i) = c_i +  
\frac{(1-c_i)}{1+\exp\{-a_i(\theta_s - b_i) \}}
\end{equation}
where $Y_{si}$ is the score of the $s$-th student to the $i$-th item, 
$\theta_s$ is the student ability parameter, $a_i$ is the item discriminant parameter, $b_i$ 
is the item difficulty parameter and $c_i$ is the item guessing parameter. Setting $a_i=1$ and
$c_i=0$ results in the common Rasch model.
\begin{figure}
 \includegraphics[width=.5\linewidth]{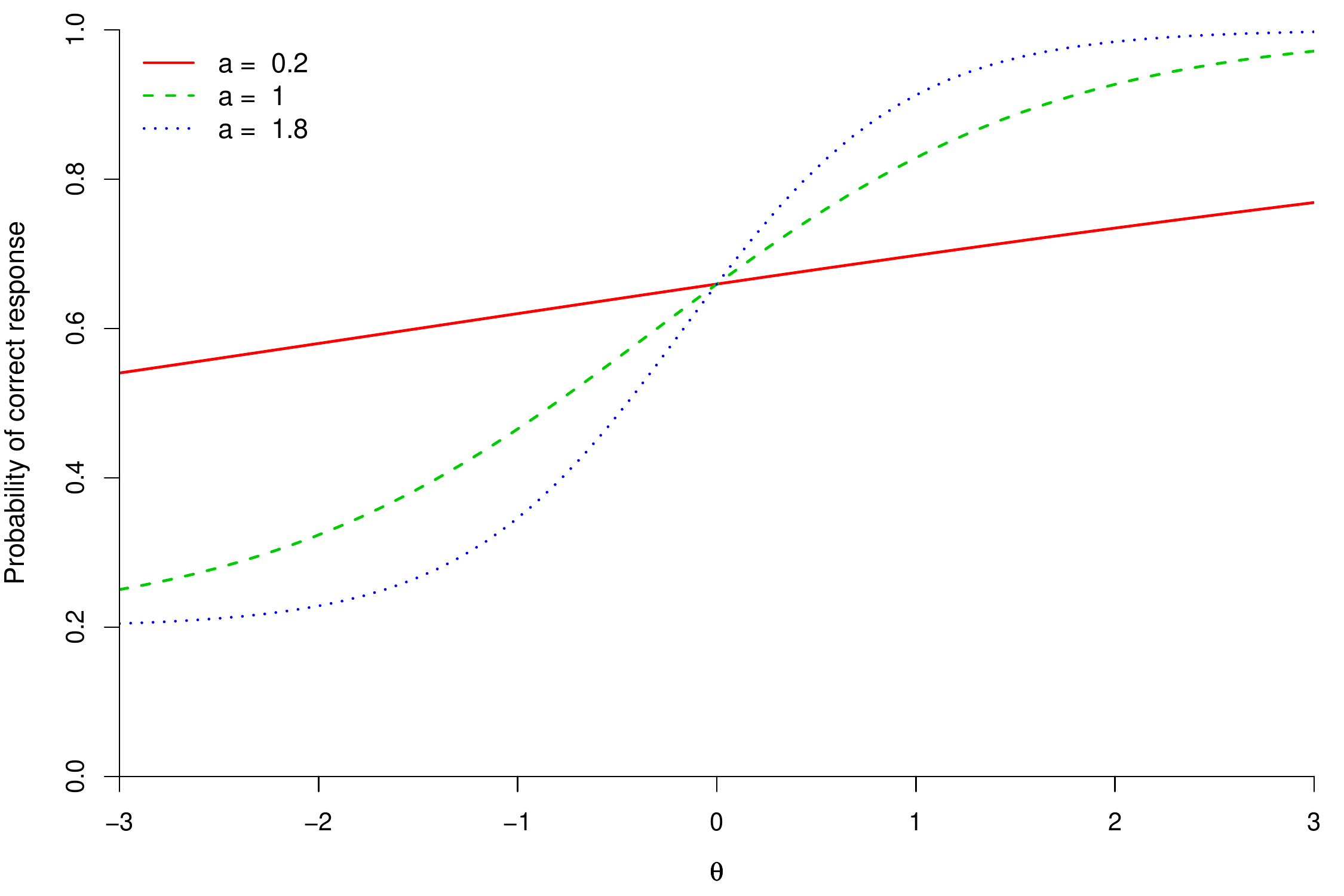}
\caption{The three-parameter logistic model with varying $a$, $b$ = 0.3 and $c$ = 0.2}
\label{fig:3pl}
\end{figure}

Common item-selectors include simple random selection and Point Fisher Information where the ``most informative'' item for this
student is chosen. Information is measured by the Fisher Information 
\[
I(\theta) = E\left[\left(\frac{\delta}{\delta \theta} \log L(\theta) \right)^2\right] 
\]
where $L(\theta)$ is the likelihood (function of ability) for fixed values of item parameters. 
This results in the item information function
\[
I_i(\theta) = \frac{P_i'(\theta)^2}{P_i(\theta)(1-P_i(\theta))}.
\]
For the three-parameter model the item information function is
\[
 I_i(\theta) = a_i^2 \cdot \frac{1-P_i(\theta)}{P_i(\theta)} \cdot \frac{(P_i(\theta) - c_i)^2}{(1-c_i)^2}.
\]

The focus in CAT \cite{wainer2000computerized} and IRT is to measure
abilities and the item selection methods were developed for that
purpose.  Since the CAT methods do not account for learning in a
dynamic environment, new models and item selectors needs to be
developed for the purpose of increased learning. The need for this
becomes apparent as models are fitted to actual data, as seen 
below.

\section{The tutor-web}
\begin{figure}[!h]
\centering
\rotatebox{-90}{\includegraphics[width=0.5\linewidth]{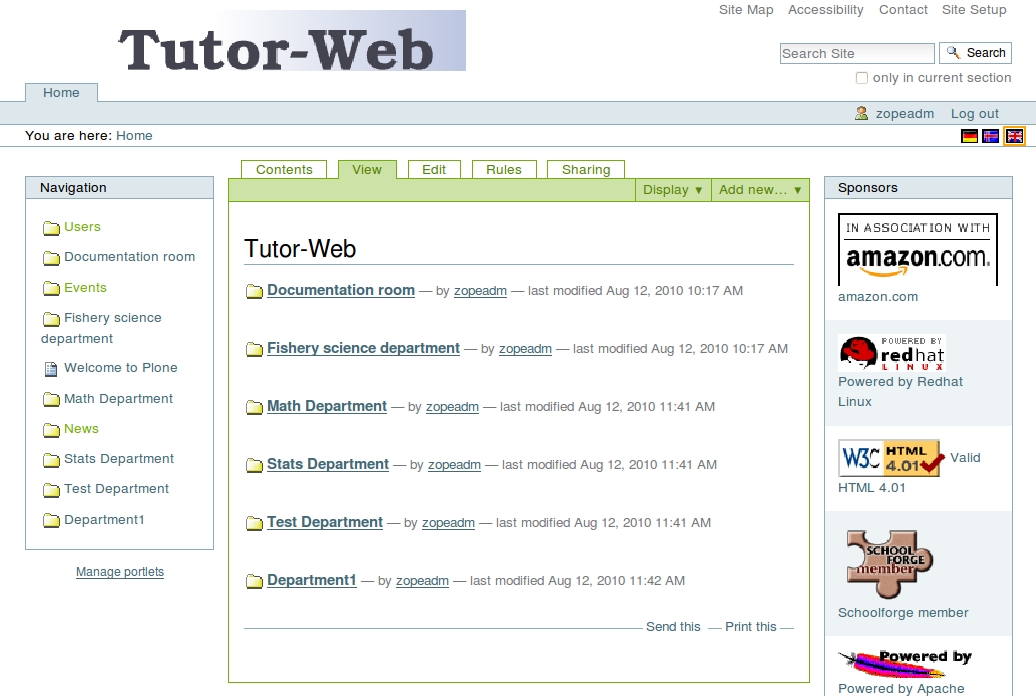}}
\caption{The tutor-web main page.}
\end{figure}

The tutor-web, used here,
is a freely accessible web-based university which has
been used for computer-assisted education in mathematics and
statistics and research on education. Needless to say, the
tutor-web is not the only such system.
%
%
%
With the increasing number of web-based educational
systems several
types of educational
systems have emerged. These include learning management system (LMS),
learning content management system (LCMS), virtual learning
environment (VLE), course management system (CMS) and Adaptive and
intelligent Web-based educational systems (AIWBES).\footnote{
The terms VLE and
CMS are often used interchangeably, CMS being more common in the
United States and VLE in Europe.}
The LMS is designed for planning, delivering and managing
learning events, usually adding little value
to the learning process nor  supporting internal content
processes \cite{ismail2001design}. A VLE  provides similar
service, adding interaction with users and access to a
wider range of resources \cite{piccoli2001web}. The primary role of a
LCMS is to provide a collaborative authoring environment for creating
and maintaining learning content \cite{ismail2001design}.

\begin{figure}[!h]
\centering
\includegraphics[width=.5\linewidth]{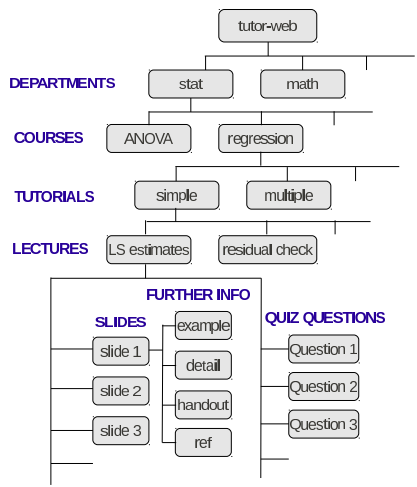}
\caption{The structure of the tutor-web}
\label{fig:layout}
\end{figure}

Many systems are merely a network of static hypertext pages
\cite{brusilovsky1999adaptive} but adaptive and intelligent Web-based
educational systems (AIWBES) use a model of the goals, preferences and knowledge of each student and
use this to adapt to the needs of that student
\cite{brusilovsky2003adaptive}. These systems tend to be subject-specific because of their structural complexity and therefore do not provide a broad range of content.

The tutor-web (at http://tutor-web.net) 
is an open and freely accessible \mbox{AIWBES} system, available
to students and instructors at no cost. The system has been a research
project since 1999 and is completely based on open source computer
code with material under the Creative Commons Attribution-ShareAlike
License. The material and programs have been mainly developed in
Iceland but also used in low-income areas (e.g. Kenya).  Software is
written in the Plone\footnote{http://plone.org}, CMS (content
management system), on top of a
Zope\footnote{http://zodb.org} Application Server.

In terms of internal structure, the material is modular, consisting of
departments (e.g. math/stats), each of which contains courses
(e.g. introductory calculus/regression). A course can be split into
tutorials (e.g. differentiation/integration), which again consist of
lectures (e.g. basics of differentiation/chain rule).  Slides reside
within lectures and may include attached material (examples, more
detail, complete handouts etc). Also within the lectures are drills,
which consist of multiple-choice items.  These drills/quizzes are
designed for learning, not just simple testing.  The system has been
used for introductory statistics\citep{stefansson2004twe},
mathematical statistics, earth sciences, fishery science, linear
algebra and calculus\citep{2013arXiv1310.8236J} in Iceland and 
Kenya\footnote{http://tutor-web.tumblr.com/post/59494811247/web-assisted-education-in-kenya}, with some 2000 users
to date. 

The whole system is based on open source software and the
teaching material is licensed under the Creative Commons
Attribution-ShareAlike License\footnote{http://creativecommons.org/}.
The system offers a unique way to
structure and link together teaching material. The structure of the
tutor-web can be seen in Fig. \ref{fig:layout}.

An important part of the system are the interactive drills where the
emphasis is on learning rather than evaluation. A student can continue
requesting drill items until a satisfactory grade is obtained. The
grade is currently calculated as the average of the last 8 questions
per lecture in the current version of the system, but alternatives are
considered below.

\section{Analyses and results}
\subsection{Some experimental results}

\begin{figure}
 \includegraphics[width=.75\textwidth]{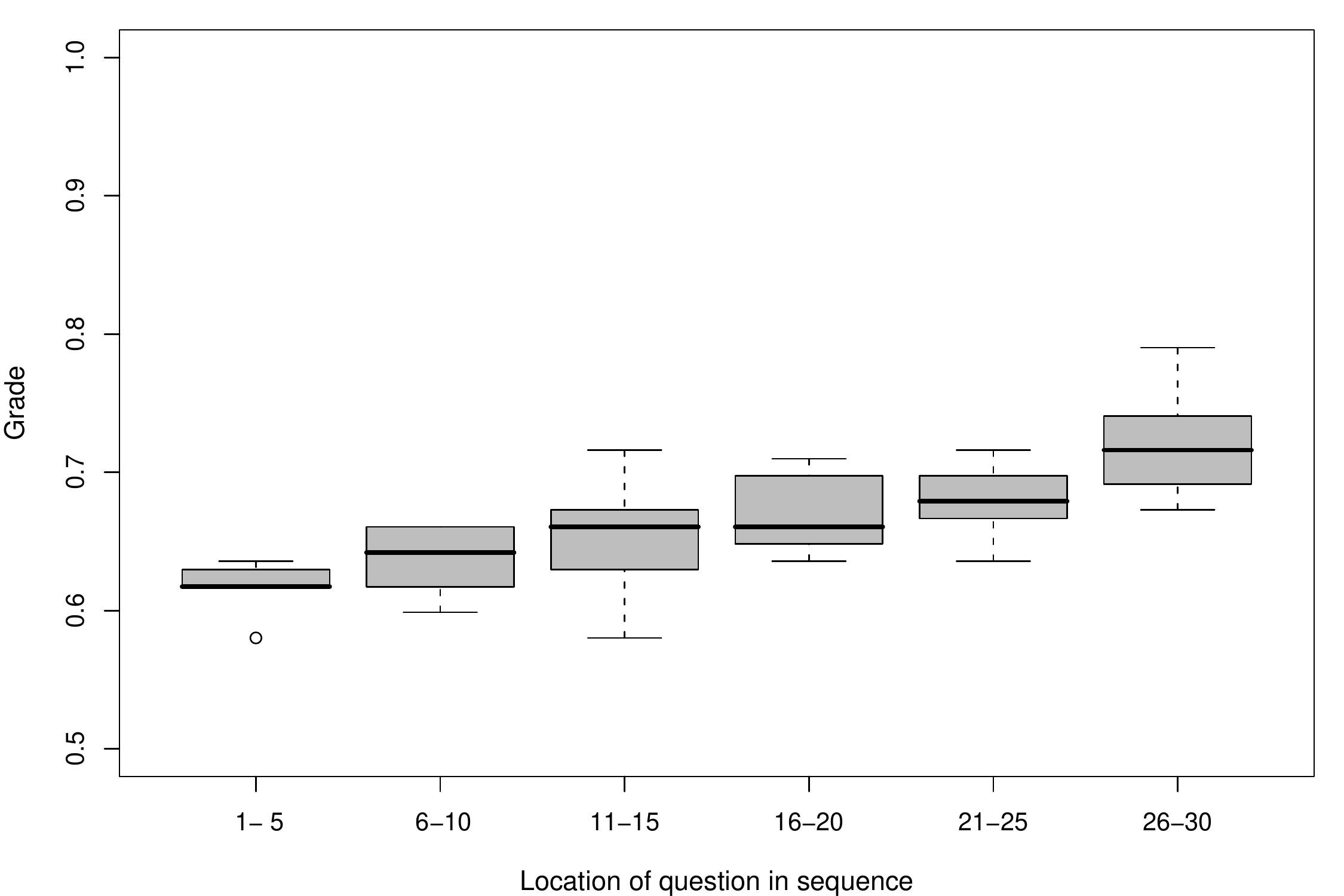}
\caption{Grade development based on averages across 162 students in an
introductory statistics course.}
\label{fig:means}
\end{figure}

The most important part of the tutor-web is a drilling system, the whole point
of which is to induce learning, rather than evaluation.  It is seen in
Fig. \ref{fig:means} that the mean grade to a question increases as
the students see more questions, as is to be expected.  Although this
does not automatically imply understanding, it does imply that student
knowledge changes during the use of the system.  From this it follows
that the usual IRT models do not apply to the present setup, nor do
any conceptual models or frameworks designed only for testing purposes.

\subsection{Model results}
The final fitted model, based on retaining statistically significant
variables becomes:
\begin{equation}
\label{eqn:fit}
\begin{split}
\log\left( \frac{p}{1-p} \right) & = \beta_1 \cdot \texttt{rankdiff} +
\beta_2 \cdot \texttt{numseen}  + \beta_3 \cdot \texttt{numseen}^2 + \\ &
\beta_4 \cdot \texttt{numseen}^3 + \beta_5 \cdot \texttt{natt} +
\beta_6 \cdot \texttt{natt}^2 +  \texttt{sid} ,
\end{split}
\end{equation}
where \texttt{rankdiff} is the ranked difficulty of the question (at
the time of analysis), \texttt{numseen} is the number of times the
particular question has been answered (seen) by the student, \texttt{natt} is
the total number of attempts the student has made at questions in the
lecture, and \texttt{sid} is the student id.

\begin{figure}
 \includegraphics[width=.75\linewidth]{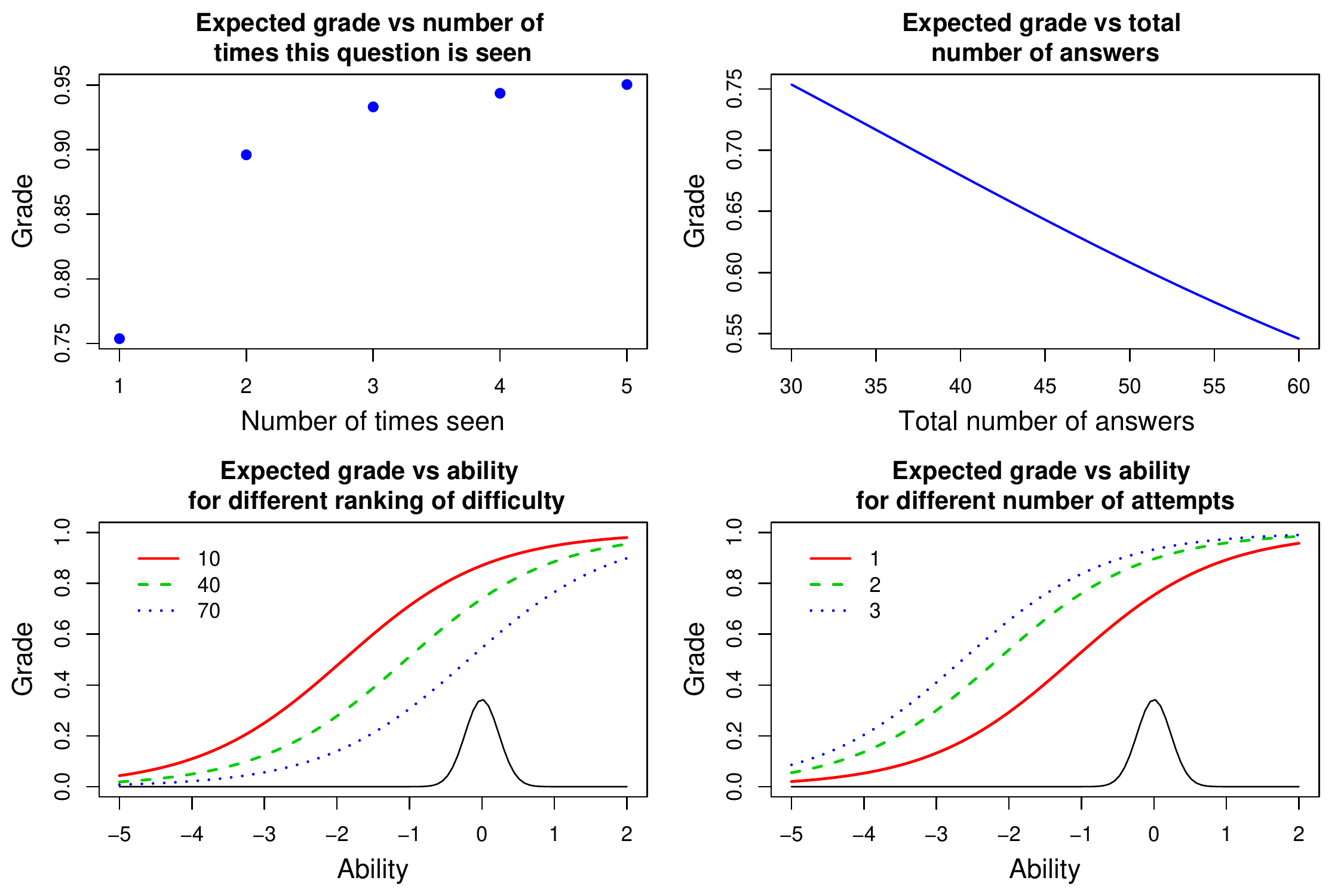}
\caption{Top panels: Model predictions of average grade as a function
  of (a) the number of times a question is seen and (b) the total
  number of answers given. Bottom panels: (c) Expected grade as a
  function of ability and (d) Expected grade as a function of ability,
  for different numbers of attempts at the question.  The density
  shown in the lower panels indicates the distributions of estimated
  student ability.}
\label{fig:glmm}
\end{figure}

It should be noted that the fitted equation \ref{eqn:fit} is quite
different from the usual IRT equation \ref{eqn:3pl} since the fitted
equation needs to takes learning into account. This is done by
explicitly stating the number of times an item has been seen as well
as the number of questions requested. 

The two-parameter Rach model only incorporates the first and last
terms in equation \ref{eqn:fit}, but the remaining parameters are also
statistically significant and therefore also needed to explain the
data at hand.



\section{Discussion}
It is seen that the specific question is not a significant contributor
to explaining the data once question difficulty has been inserted as a
parameter.

If the student answers a great many times, then the model predicts a
lower grade, consistent with assuming this corresponds to
guessing. This result will almost certainly depend on the range of
attempts made available to the model. Thus, since the current data
include up to 60 attempts within a single lecture, the upper end will
almost certainly correspond to guessers. Within this particular course
a minimal return rate of 20 correct answers per lecture was required,
and each such on-line lecture contained items corresponding to 1-2
weeks of material in the actual course.  This is quite different from
a system with smaller on-line lectures and lower return requirements
from each lecture. In such a course one would expect to see a true
quadratic effect with a maximum grade at an intermediate number of
attempts, given enough data.

If the student sees the same question more than once then there is clear
evidence of learning.

\subsection{General experience}
The tutor-web has been used for teaching mathematics and statistics to
students of mathematics, biology, geography and tourism with
considerable student satisfaction \citep{stefansson2004twe}. In a recent survey of 60
students using the tutor-web, 53 of the students indicated that they
had less difficulty answering the last questions in a drill session
than the first and all of the students answered ``yes'' to the
question ``I learn by taking quizzes in the tutor-web,'' in accordance
with the numerical results from this study.

\subsection{Current work}

It is reasonably clear that an item allocation method which simply
hands out items with equal probability is not optimal. In particular
this does not guarantee that a beginner first sees ``easy'' items nor
that a student who completes the lecture or course has had to answer
the most difficult items as a part of the way towards a high grade.

\begin{figure}[!h]
\includegraphics[width=.75\linewidth]{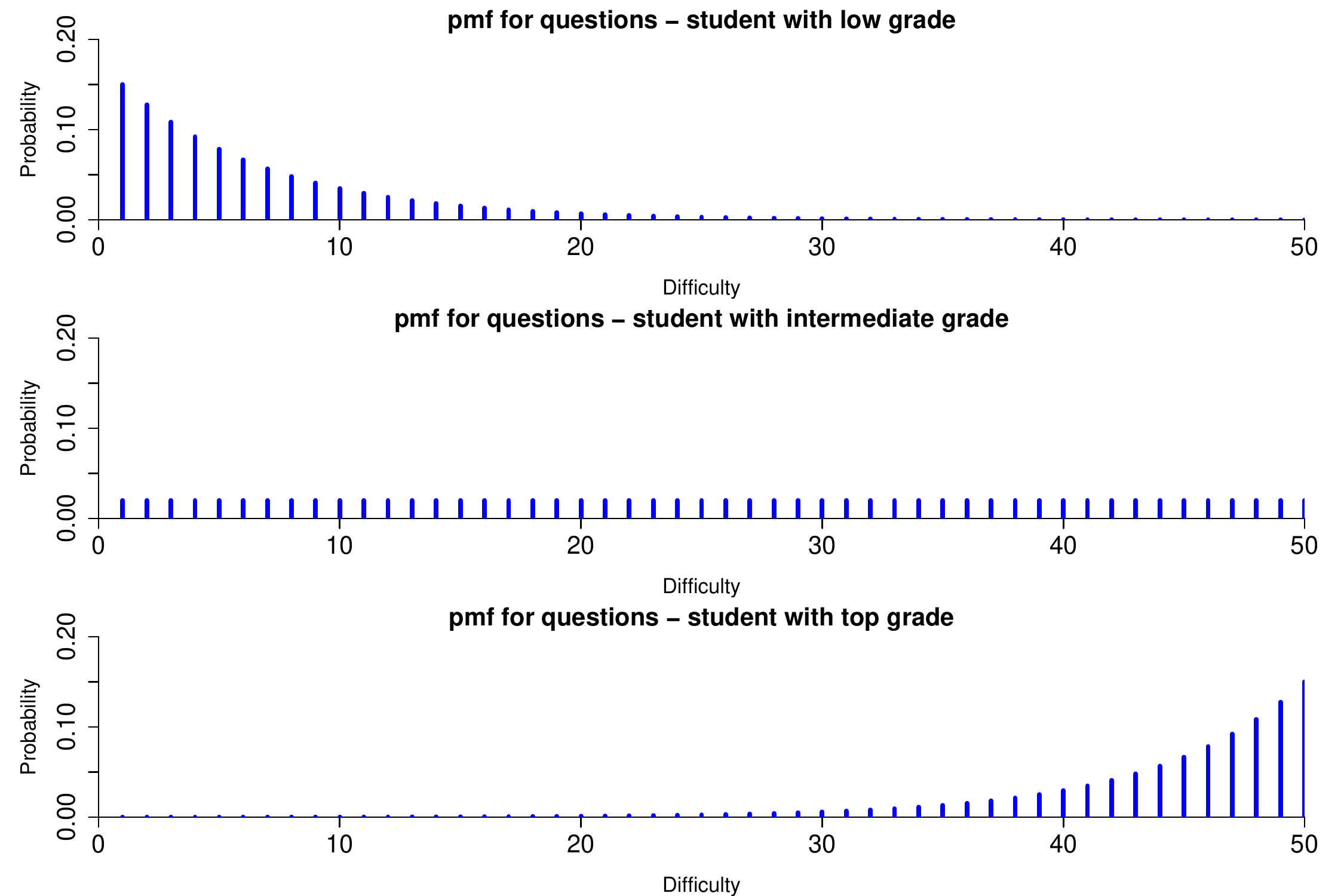}
\caption{Possible development of pmf for questions as grade develops.}
\label{fig:pmf}
\end{figure}

Current development is therefore focused on implementing the
following item allocation rules within the system:
\begin{itemize}
 \item select easy questions in the beginning
 \item increase item difficulty as the  grade  increases
 \item select, with some probability, questions the student has done incorrectly in the past  
 \item select, with some probability, questions from old material to refresh memory.
\end{itemize}

Fig. \ref{fig:pmf} gives one implemented development of the
probability mass function (pmf), as a function of item difficulty.
The panels indicate how the pmf varies as a function of the student's
grade. This simple approach alone implements a personalized approach
to the drills and it satisfies several important criteria, first by
starting with easy items and second by ensuring that a student does
not leave with a high grade without answering difficult items.

\subsection{Item database design}
Consider for the moment the 3PL in Fig. \ref{fig:3pl}. One aspect
of item design is to design items which classify students. Each should
therefore have a steep slope and they should have different
midpoints. For the data considered here, however, Fig. \ref{fig:glmm}
indicate that the student ability distribution lies quite far to the
right on the scale, when compared to the difficulty of easy items, but
matches the mid-point of the most difficult items considered here.
If the ability was static and the items were drawn at random one would
only conclude that a batch of more difficult items is needed, but the
dynamic nature of the on-line study is more complex.

The last panel of Fig. \ref{fig:glmm} demonstrates how the
distribution of ability is too far to the right compared to the mean
difficulty of items recieved in the first attempt within a
lecture. However, as the number of attempts increase, the student's
ability increases (panel a) and this leads to an upwards shift in the
expected grade as see in panel d.

What is needed is a combination of several approaches: The item
allocation algorithm needs to take into account both the item
difficulty level and link this to the student's dynamic ability. The
simples such approach is merely to increase the mean difficulty as the
grade increases, as is done in Fig. \ref{fig:pmf}, but this is not
enough as is seen in panel d of Fig. \ref{fig:glmm} : (i) First, the
item design needs to ensure that even the best students will, at the
peak of their learning, still recieve difficult items, as measured on
their personalized scale. These items are much more difficult than
could possibly be administered randomly to a random group of students.
(ii) Second, the link between the mean and ability is not quite
trivial, also as seen in the same panel: In this IAA the mean item
difficulty is linearly linked to the student's grade and this lifts
the mean grade too much compared to the ability distribution. A
different link could be chosen to ensure that students get items
which, at every ability and learning level always have a probability
of a correct answer closer to 0.5.

Next, the grading scheme itself can be modified in many ways.  In
the present setup the average grade for the previous 8 answers is used
as a ``lecture grade''. Again, many alternate approaches could be
designed in order to entice the student to continue. These include a
longer or expanding tail (i.e. more than 8, e.g. $\max(8,n/2)$ where
$n$ is the number of attempts) and/or tapering, where the most recent
answers get more weight in the average.

Finally, timeout options do not exist within the tutor-web. It would
be an interesting experiment to investigate how different timeout 
functions affect behavior.

Overall, it is seen that student behavior and corresponding model
results are quite different for the on-line student in a learning
environment, as compared to a student in a testing environment. This
leads to new considerations and research on how these dynamic
environments can be designed so as to maximize student learning as
opposed to just estimating student knowledge with a high degree of
precision.

\section{Acknowledgements}
The tutor-web project has been supported by 
the University of Iceland, 
the United Nations University
the Icelandic Ministry of Education and
the Marine Research Institute in Reykjavik. 
Numerous authors have contributed educational material to
the system. The system is based on Plone
\footnote{http://www.plone.org} and educational material is mainly
written in \LaTeX.  Examples and plots are mostly driven by R
\citep{rcore}.  The current Plone version of the tutor-web has been
implemented by Audbjorg Jakobsdottir but numerous computer programmers
have contributed pieces of code during the lifetime of the project.

Preliminary analyses in this work were first presented at EduLearn11.

%
%


\end{document}